\documentclass[12pt]{article}
\usepackage{epsfig}
\usepackage{graphicx}

\def\beq{\begin{equation}}
\def\eeq#1{\label{#1}\end{equation}}
\def\eeqn{\end{equation}}

\def\beqa{\begin{eqnarray}}
\def\eeqa#1{\label{#1}\end{eqnarray}}
\def\eeqan{\end{eqnarray}}

\let\bar=\overbar

\def\Dslash{\not{\hbox{\kern-4pt $D$}}}
\def\dslash{\not{\hbox{\kern-2pt $\del$}}}

\def\msb{{\bar{\ssstyle M \kern -1pt S}}}

\usepackage{fancyhdr,graphicx}
\def\Title#1{\begin{center} {\Large {\bf #1} } \end{center}}

\begin{document}

\Title{New Results on the Time lags of the Quasi-Periodic Oscillations in the Low-mass X-ray Binary 4U 1636--53}

\bigskip\bigskip


\begin{raggedright}

{\it 
Marcio G B de Avellar$^{1}$~~Mariano M\'endez$^{2}$~~Diego Altamirano$^{3}$~~Andrea Sanna$^{4}$~~and Guobao Zhang$^{5}$\\
\bigskip
$^{1}$Instituto de Astronomia, Geof\'isica e de Ci\^encias Atmosf\'ericas, Universidade de S\~ao Paulo, Rua do Mat\~ao 1226, 05508-090, S\~ao Paulo, Brazil, mgb.avellar@iag.usp.br\\
\bigskip
$^{2}$Kapteyn Astronomical Institute, University of Groningen, P.O. Box 800, 9700 AV Groningen,
The Netherlands, mariano@astro.rug.nl\\
\bigskip
$^{3}$School of Physics and Astronomy, University of Southampton, Southampton, SO17 1BJ,
United Kingdom, d.altamirano@soton.ac.uk\\
\bigskip
$^{4}$Dipartimento di Fisica, Universit\`a degli Studi di Cagliari, SP Monserrato-Sestu km 0.7, I-09042, Monserrato,
Italy, andrea.sanna@dsf.unica.it \\
\bigskip
$^{5}$New York University,
Abu Dhabi, UAE, zhanggb.astro@gmail.com\\

}

\end{raggedright}

\section{Introduction}

\begin{figure}[htb]
\centering
\includegraphics[width=0.45\textwidth,height=0.40\textwidth,angle=0]{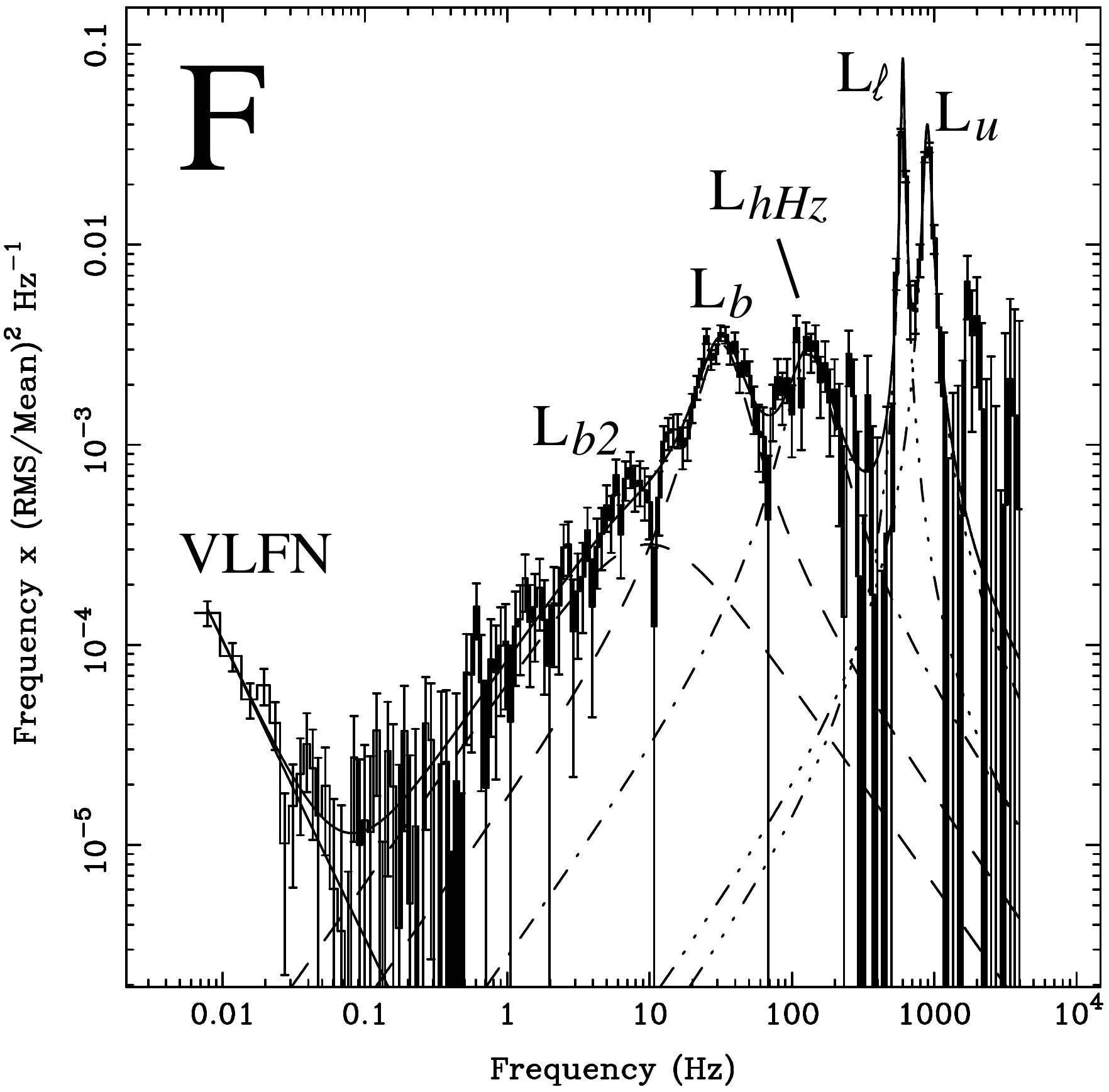}
\includegraphics[width=0.45\textwidth,height=0.42\textwidth,angle=0]{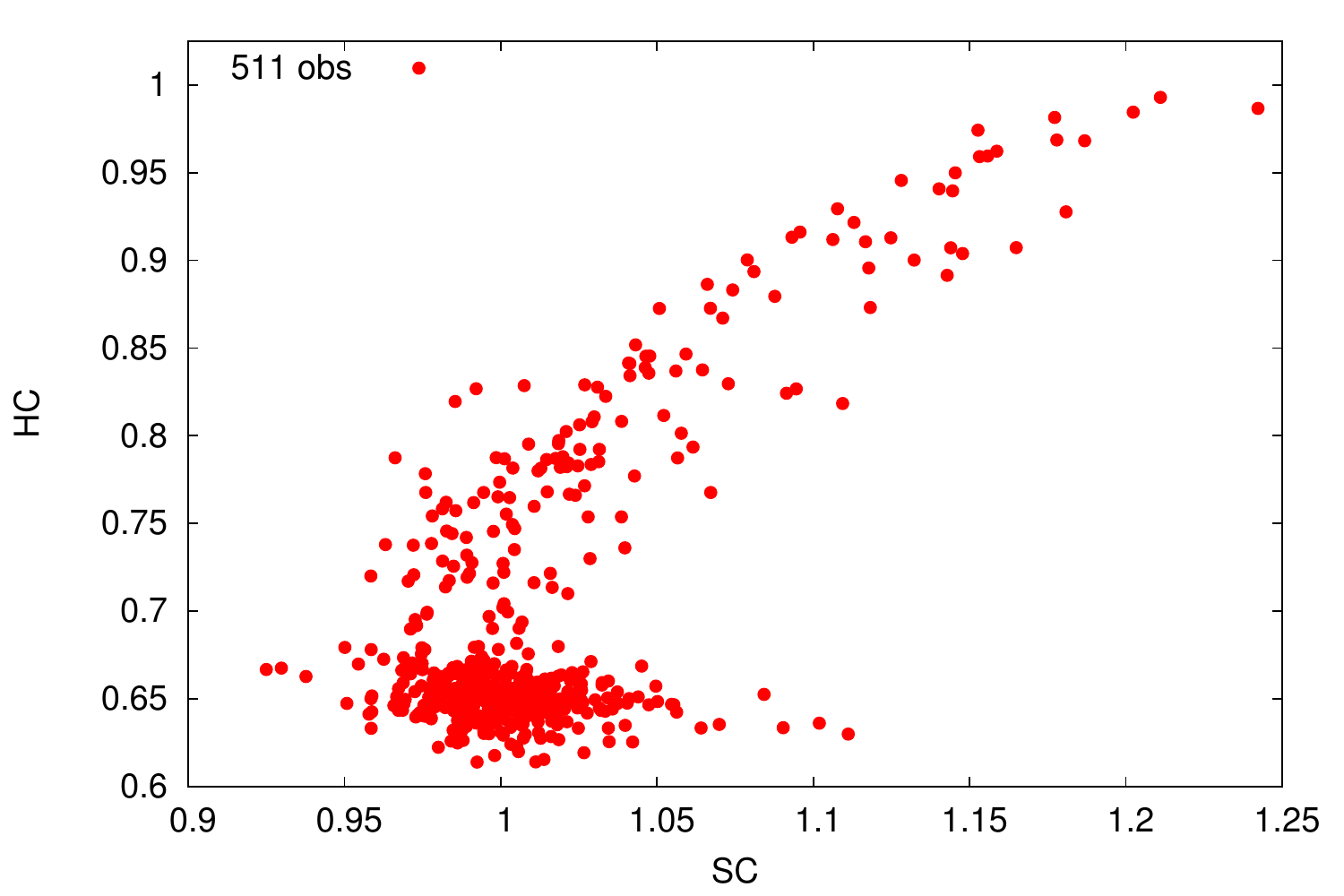}
\caption{Left panel: the variability features at a gives source state. Right panel: hard X-ray colour {\it vs} soft X-ray colour for a subset of 511 observations of 4U 1636--36.}
\label{fig01}
\end{figure}

The astrophysical system 4U 1636--53 is binary stellar system where a neutron star accretes matter from ordinary low-mass star ($M\sim 0.4~M_{\odot}$, hence the classification ``low mass X-ray binary'') primarily via direct Roche-Lobe overflow. Such systems show a myriad of variability features in the X-ray light curve, ranging from mili-hertz to kilo-hertz, and it is now well known that most of them, identified as quasi-periodic oscillations (QPOs), correlate with each other in a given source and among sources of the same type, and that their appearance depend on the spectral state of the source, e.g., on its position in the colour-colour diagram (see [1] and [2]). See Figure \ref{fig01}.
 
Most of the work in the last 15 years have focused in the characterization and properties of these variability features, e.g., in [1], [2], [3] and [4]; or the correlations between the fastest variability feature, the kHz QPOs, and other typical high and low frequencies features. In particular for the kHz QPO, the idea that they can reflect the inner radius of the accretion disc has received much attention due to its importance for the physics of matter under extreme regimes, e.g., [5] and [6]. Models for the kHz QPOs (e.g., [5] and [6]) assume clumps of matter orbiting the star at a preferred radius in the disk, something that could constrain the EOS. The finding that the frequency $\nu_{hHz}$ of the hecto-Hertz QPO $\sim$ constant leads to diskoseismic models for them like in [7] where $\nu_{hHz}$ $\Leftrightarrow$ orbital frequency at the radius where a warped disk is forced to the equatorial plane by the Bardeen $\&$ Petterson effect; and like in [8] where Rayleigh-Taylor gravity waves at the disk-star boundary layer originate the hHz QPOs.
 
Much less work has been done on another aspect of the X-ray signal: The energy- and/or frequency-dependent time (or phase) lags. These are Fourier-frequency-dependent measurements of, respectively, the time (phase) delay between two concurrent and correlated signals, in this case light curves of the same source, in two different energy bands (see [9] for a deep explanation). Time lags generally appear in systems with accretion, and the proposed mechanisms usually involve Compton scattering or reflection off the disc being tools to study the physics of accretion near the neutron star, possibly with consequences for constraining the parameters of the neutron star like the mass and the radius, since the {\it time lags} encode information about the size and geometry of the scattering medium or reflector. This can be done finding out the dependence of the {\it time lags} on the frequency of the QPO and on energy. In order to do so, we divided the data in energy bands and in frequency ranges. 
 
\section{Results}

Previous results obtained by us in [10] only for the kHz QPOs were the first study of the dependence of the time lags of both kHz QPOs upon frequency and the first on energy in the case of the upper kHz QPO. The main findings were: a) Regarding the frequency dependence $\Rightarrow$ $\nu_{low1636}$: $\bar{\Delta t}=-21.0 \pm 0.6 \mu s$;$\nu_{upp1636}$: $\bar{\Delta t}=+11 \pm 3 \mu s$; and b) Regarding the energy dependence $\Rightarrow$ $\nu_{low1636}$: Decay=$3.6 \pm 0.3 \mu s/keV$; $\nu_{upp1636}$ $\bar{\Delta t}=+4 \pm 3 \mu s$. We clearly see the inconsistency between the lags of the lower and the upper kHz QPOs; light travel time arguments imply that the region where the lags are produced should be $\sim 3-30km$; the small variation of the lags with frequency implies that the geometry varies very little, $\sim 1km$. The model presented in [11] can explain our results via inverse Comptonization (for some specific parameters of the system) solving the linearised time dependent Kompaneets equation.

\begin{figure}[htb]
\centering
\includegraphics[width=0.45\textwidth,height=0.40\textwidth,angle=0]{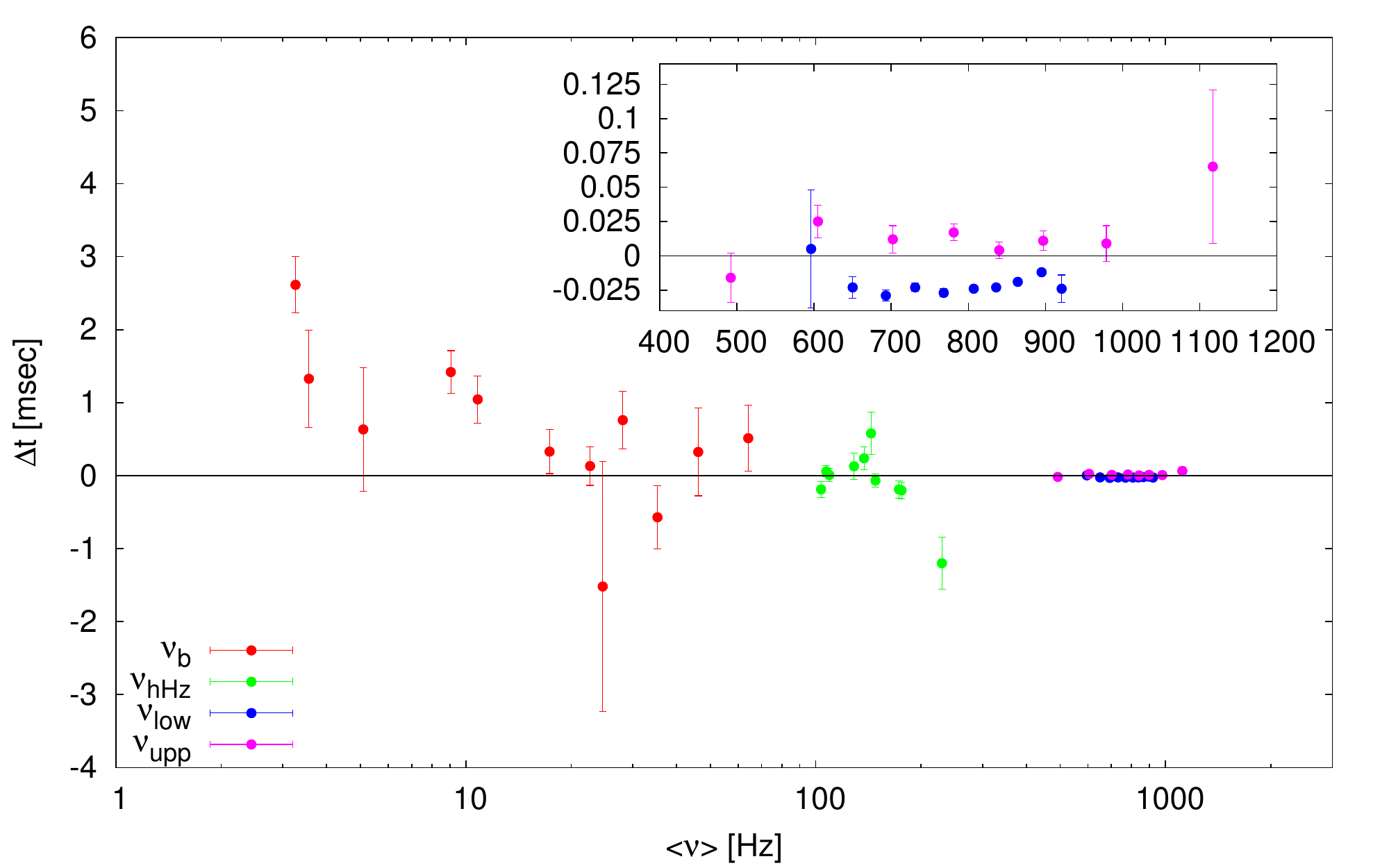}
\includegraphics[width=0.45\textwidth,height=0.40\textwidth,angle=0]{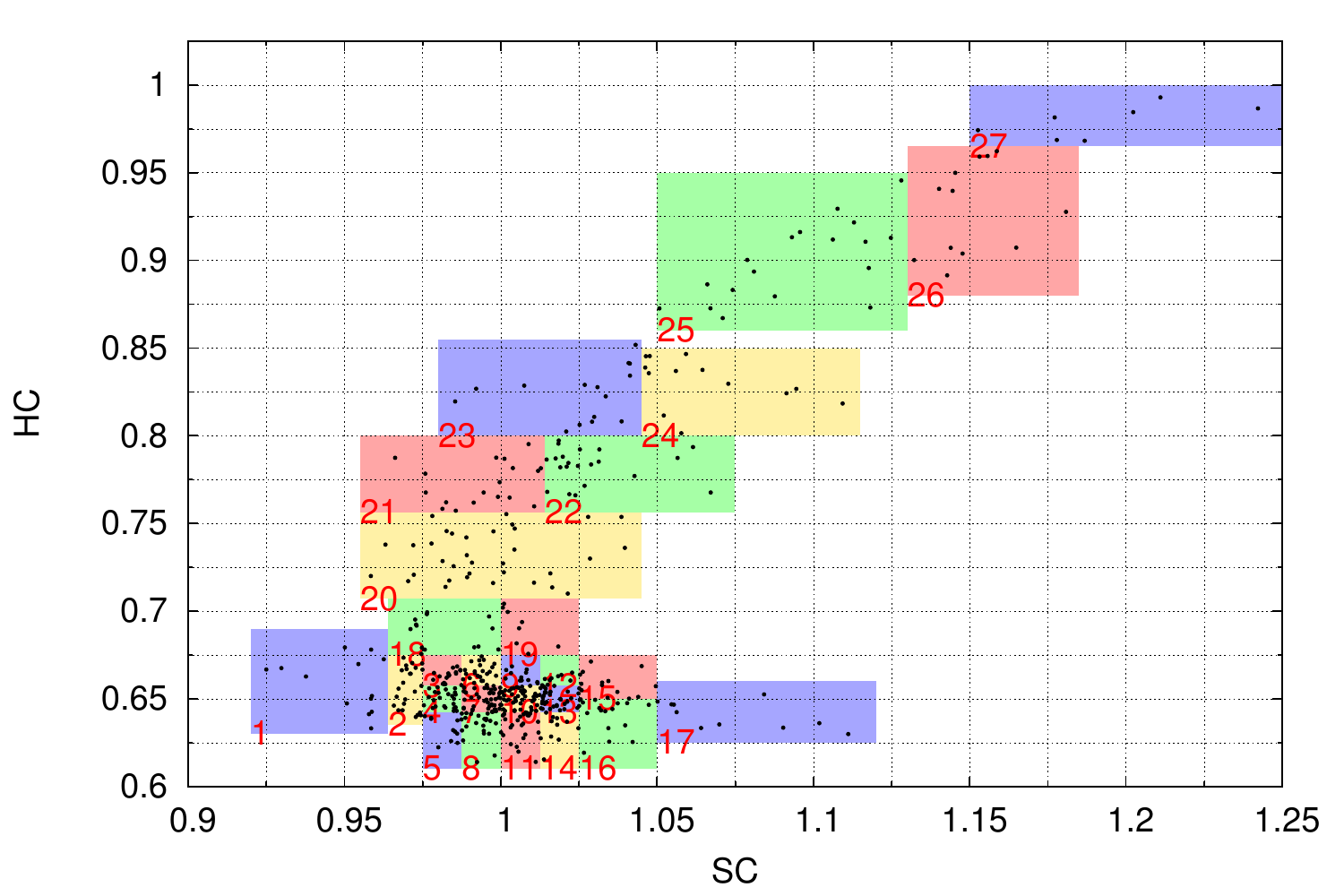}
\caption{Left panel: Time lags {\it vs} QPO frequency for all QPOs studied. Right panel: Inclusion of the remaining data set; new methodology; possibility of studying many features and their dependence on physical observables.} 
\label{fig02}
\end{figure}

Other QPOs also play an important role in constraining models and parameters of these stellar binaries and we started a program to study their lags beginning with the lags of the hectoHertz and break frequency QPOs in a subset of 197 observations, making use of some correlations among frequencies and colours. Our results for the hHz and break frequency QPOs are shown Figure \ref{fig02}, left panel.

We found that: (i) For the break frequency QPO: for low frequencies, in general the time lag is positive, but it is decreasing with increasing frequency, reaching zero lag at $\sim$ 20 Hz. Between 20 and 35 Hz there is a small fluctuation around zero, from where the time lags become positive again and increase slightly above zero up to 65 Hz. (ii) For the hHz QPO: we see that when the frequency is $\sim$ 100 Hz the time lag is negative, but it increases to zero already at $\sim$ 110 Hz, being consistent with this value up to 130 Hz from where it increases to 0.5 msec at around 140 Hz. From 140 Hz the time lag decreases sharply, being strongly negative for hHz $>$ 220 Hz. Taking the value of the constant fit to the break frequency QPO ($+0.193~msec$) and to the hectoHertz QPO ($-0.041~msec$) as simple estimators, one concludes that the geometry of the medium varies by $58~km$ and $12~km$, respectively. However, taking seriously the frequency dependence, specially for the hHz QPO, the variation of the geometry of the scattering medium can be as big as 460 km, much bigger than for the kHz QPOs. Because there is not a tight link between the hHz QPO and the QPO at break frequency as we see between the two kHz QPOs, the model [11] cannot explain these lags and if the place where the hHz QPO is produced is nearby the location where the kHz QPOs are produced, then there must exist another mechanism operating and discriminating among the frequencies. On the other hand if Compton scattering is responsible for the lags, we need a denser medium for the hHz QPO and an even denser medium for the QPO at the break frequency in order to produce the 1--2 orders of magnitude bigger lags.

At last, because all these variability features depend on the position on the colour-colour diagram, we extended our program to include all the data that show both kHz QPOs and the others, delimiting regions on this diagram (Figure \ref{fig02}, right panel) in order to better correlate with observables. The results are just coming.

\subsection*{Acknowledgement}

We express our thanks to the organizers of the CSQCD IV conference for providing an 
excellent atmosphere which was the basis for inspiring discussions with all participants.
We have greatly benefitted from this.

\end{document}